\documentstyle[fleqn,prl,aps,preprint,epsf]{revtex}

\newcommand{\cbeta}{ { \ss }}
\newcommand{\mcl}{{\cal L}}

\newcommand{\beq}{\begin{equation}}
\newcommand{\eeq}{\end{equation}}

\begin{document}
\widetext

\title{Exact Solution of a Repulsive Fermi Model With Enhanced
Superconducting Correlations }
\author{H  R Krishnamurthy$^*$ and B  Sriram Shastry$^*$  } 
\address{Department of Physics, Indian Institute of
Science, Bangalore 560012, India} 
\maketitle

\begin{abstract}
We present the exact solution of a model of interacting
 fermions in any dimension with a pure repulsive interaction projecting
out a given Cooper channel.
The solution rests upon the 
 infinite ranged character of the interaction in real space, leading to 
a functional integral that is
dominated by a Gaussian term. The solution 
produces strong superconducting enhancements and quasi long ranged order
in a channel that is not present in the Hamiltonian explicitly, but
 of the form given  by arguments from {\em order by projection}.
\end{abstract}
\pacs{71.10.Fd, 71.10.Li, 74.20.Mn}

There is considerable recent interest \cite{pwajrs} in  the possibility
of models displaying a superconducting behaviour driven solely  by
repulsive interactions. The search is motivated by the High Tc problem, where
no obvious,known attractive interaction can account for the phenomena; so
one believes that  repulsive interactions, originating in the
Coulomb repulsion expressed within
 the Wannier basis of a few
tightbinding bands, are ultimately responsible. In one dimension, the usual
kind of repulsive interacions generically lead to enhanced SDW order, via
powerlaw correlations with small decay exponents rather than
superconducting enhancements. However recent interesting work
\cite{fabrizio} shows that under certain conditions,  superconducting
correlations of non trivial symmetry are enhanced. 
 In the physically important
 case of two dimensions the situation is not completely clear in
relation to popular models, such as the Hubbard or the t-J model.
While a treatment within  the Random Phase Approximation near
an Antiferromagnetic instability leads to d-wave superconductivity
\cite{dp}, one may worry whether the conclusions based on the (weak
coupling) approximation are valid for strong repulsions. Similar
worries exist regarding various versions of the Gauge
theories\cite{pal} that are in vogue currently. In this context, repulsive
models which can be solved exactly have an important role to play. A new
set of models was introduced recently \cite{bss1,bss2,bss3} that
demonstrate the possibility of enhanced superconducting correlations
quite  explicitly. These models are generalizations of the Hubbard model,
and include a term that is best interpreted as a
projection operator that excludes a certain Cooper pairing  channel from
the problem.  We present here an exact solution of the basic model
involving the kinetic energy and the projection operator. Our solution is
obtained by exploiting  a  certain feature of the interaction within the
framework of functional integrals: namely that 
the projection operator is an infinite ranged repulsive operator, and its
Hubbard-Stratanovic (HS) linearizing field is a single spatially uniform
dynamical mode that admits  {\em only Gaussian fluctuations in the
thermodynamic limit} .This is in contrast to models with attractive
interactions,  
where a new saddle point value of the HS field gets stabilized at low
temperatures, and its fluctuations can be ignored. The repulsive case is
much more subtle, and has some similarities to the situation found in 
Fermionic models in infinite dimensions \cite{metzvoll} where the relevant
HS field is a Grassman variable that factorises the kinetic energy,  and
has only Gaussian fluctuations \cite{fnotehrk}. We find that as a
consequence the  fermi gas develops  enhanced correlations in a 
``compromise'' pairing channel which is not explicitly present in the
Hamiltonian.  These are precisely of the sort that is expected from
arguments from {\em order by projection} \cite{bss1,bss2} . The earlier
treatments \cite{bss1,bss2} have used a variety of non perturbative
techniques, such as rigorous inequalities and variational approaches. The
exact solution presented here is consistent with these, and give us in
addition, a clear understanding of the origin of these enhancements and
of the quasi LRO in terms of the singularities of the two particle
scattering amplitude. 
The model is defined by the Hamiltonian 
\beq
 H=T+U_{s}\,B^{\dag }B \;\; . \label{hamiltonian}
\eeq
Here $ B \equiv \sum \zeta(k)\; b_{k}$
is a Cooper pair operator,  $
b_k = c_{-k\downarrow }c_{k\uparrow }$ are the pair destruction operators, 
$T=\sum \epsilon _{k}n_{k\sigma }$ is the kinetic energy corresponding to 
a band dispersion $\epsilon _{k} [ = -2\sum_{\alpha =1}^{d}  \cos
(k_{\alpha })$ in the nearest neighbour problem~].  
$\zeta(k)$  may be chosen arbitrarily. The two cases of 
interest in 2-dimensions
are (i)  $\zeta(k)=1$ giving rise to extended s-wave order, and (ii)  
$\zeta(k) = \cos(k_x) - \cos(k_y)$ giving rise to second-neighbour d-wave
order. For simplicity of presentation we initially focus on $\zeta=1$ and
return to the other case later.

With the above choice of the model,
 we note that $B$ may be rewritten as
$\sum_r c_{\downarrow}(r) c_{\uparrow}(r)$, and hence the interaction term
may be viewed as an infinite ranged hopping term for doubly occupied
sites.

Using the H-S linearization within the  standard
Grassman variable\cite{grass} formulation, we write  the partition
function for this model as the functional integral $ Z = \int D\, c^*
D\, c \; \; D \phi^* \; D\phi\;   \exp{ - \beta \Psi }$. The free energy
functional $\Psi$ is given  in terms of the fermi fields $c, c^*$ and the 
auxiliary bose fields $ \phi, \phi^* $ as 
\begin{eqnarray} \beta \Psi & = & \int_0^\beta \phi^*(\tau)
\phi(\tau) \; d\tau - \sum_{k,\sigma} \int_0^\beta c^*_{k,\sigma}(\tau)
(\partial_\tau - \xi_k)  c_{k,\sigma}(\tau) \; d\tau   \nonumber \\
& & -   i \sqrt{U_s} \int_0^\beta
 (\phi^*(\tau) B(\tau) + B^*(\tau) \phi(\tau) )\; d\tau, 
\end{eqnarray}
where $\xi_k= \epsilon_k - \mu$. 
We use a Fourier series expansion 
$c_{k,\sigma}(\tau)= \sum_n \exp(i \omega_n \tau) \hat{c}_{k,\sigma}(n)$,
and $\phi(\tau) = \sum_n \exp( i \Omega_n \tau) {\hat{\phi}}_n$, where the
fermionic frequencies $\omega_n = ( 2 n +1) \pi \beta^{-1}$ and the
bosonic frequencies $\Omega_n = 2 n  \pi \beta^{-1}$. We denote the
Fourier components  $ \hat{b}_k(n) = \sum_m \hat{c}_{-k, \downarrow}(m)
\hat{c}_{k,\uparrow}(n-m)$ and so define ${\hat{B}}_n = \sum_k
\hat{b}_k(n-1)$.  In terms of these we can rewrite
$$\Psi = \sum {\hat{\phi}^*}_n {\hat{\phi}}_n + \sum_{k,n}(\xi_k -
i\omega_n )  \hat{c}^*_{k,\sigma}(n)\hat{c}_{k,\sigma}(n) + i \sqrt{U_s}
\sum_n( {\hat{\phi}^*}_n  {\hat{B}}_n + {\hat{\phi}}_n {\hat{B}^*}_n ).$$

We  now trace out the Fermi degrees of freedom and find the reduced free
energy functional $\Psi_\phi= \Psi_0 +\sum {\hat{\phi}^*}_n 
{\hat{\phi}}_n  - k_b T \sum_k Tr\log \{ 1 - C(k)\} $. Here  $\Psi_0 $ is
the noninteracting  free energy,   $C$ is an infinite
dimentional matrix  defined  by its (frequency space) matrix elements
$C_{n,m}(k)= U_s \sum_l G_0(k,n) G_0(k,m) {\hat{\phi}^*}_{m+l+1}
{\hat{\phi}}_{n+l+1} $  in terms of the free Green's function
$G_0(k,n)=1/(i \omega_n - \xi_k) $ , and the trace is in the frequency
space.

We  now analyse $\Psi_\phi $ in detail. We can expand 
\begin{eqnarray}
\Psi_\phi & = & \Psi_0+   \sum_n  \{  1 + U_s \mcl  {\Pi_0}(i\Omega_n)\} 
{\hat{\phi}^*}_n {\hat{\phi}}_n   \nonumber \\
& & + \frac{ U_s^2}{2} \mcl \sum \Gamma_{(m_1,m_2,n_1,n_2)}
{\hat{\phi}^*}_{m_1} {\hat{\phi}^*}_{m_2} {\hat{\phi}}_{n_1} 
{\hat{\phi}}_{n_2} + O(U_s^3). \label{expansion} \nonumber  
\end{eqnarray}
Here $ {\Pi_0}(i\Omega_n) 
 \equiv \frac{1}{\mcl} \sum_k {\pi_{k,n}} $
where $\pi_{k,n}$ are the polarizations
\beq
\pi_{k,n} 
= (2 f(k) - 1)/(i\Omega_n - 2 \xi_k) \; , \nonumber 
\eeq
and $f(k)$ is the usual noninteracting Fermi function. The fourth order
term is given explicitly as 

$$\Gamma_{(m_1,m_2,n_1,n_2)} = \frac {1}{\mcl}\sum_k {\frac   {(2 f(k)-1)
(D_{m_1,k}+ D_{m_2,k}) \delta_{m_1+m_2,n_1+n_2}  }
{D_{m_1,k} D_{m_2,k} D_{n_1,k} D_{n_2,k} } },  $$
where $D_{m,k} \equiv  (i\Omega_m - 2 \xi_k)$ .        

We note that both  $\Pi_0$ and $\Gamma$ in the above equations are of
$O(1)$ since these are normalized sums over momenta. Indeed every term in
the expansion  has a similar structure and is of the same order, namely
$O(\mcl)$. Hence one has  the remarkable  exact result that {\em the
Gaussian term dominates the rest of the terms   in the thermodynamic
limit} \cite{posit}.  Roughly speaking, the Gaussian
piece gives us the typical size ${\hat{\phi}}_n \sim
\frac{1}{\sqrt{\mcl}}$, and so the quadratic ( in $U_s$)  piece is of the
order $1/{\mcl}$, and likewise the $m^{th}$ term is of the order
$1/\mcl^{(m-1)}$.  Thus in the thermodynamic limit, it suffices to keep
the Gaussian term and to drop the remaining  terms.   
This leads to the following remarkably simple result \cite{fnotecorr} 
\beq
< {\hat{\phi}^*}_n {\hat{\phi}}_n > = \frac{\beta^{-1}}{1 +  U_s
\mcl {\Pi_0}(i\Omega_n)} . \label{basic} 
\eeq
From the same arguments, the correlation function of
the Cooper pair operators is given by
\beq
<b^*_k(n) b_{k'}(n)> \; = \;\delta_{k,k'} \; \pi_{k,n} - \frac{1}{\mcl}
\; \pi_{k,n} V_{eff} (i\Omega_n)\pi_{k',n} \label{main}
\eeq
where the  effective interaction \cite{fnotecorr} 
\beq
V_{eff}(n)= \frac{V}{1 + V {\Pi_0}(i\Omega_n)} \label{veffective}
\eeq
in terms of the (very large) coupling constant $V= U_s \mcl$.

We next present an alternate
derivation of the above results starting from the equations of motion,
which gives some more insight into them.

Define (as in ref \cite {bss2}) the set of operators  $I_l \equiv  \sum
(\epsilon _{k})^l b_k $, and $T_l \equiv \sum (\epsilon_{k})^l
(n_{-k\downarrow }+ n_{k\uparrow} -1 ) $ . Clearly $I_0 = B $ and $ T_1 =
T $.  It is easy to ascertain that 
\beq
 [I_l , T] = 2 I_{l+1} \;, \;\; [I_l , \hat{N}] = 2 I_l \;\; ,\;\;
[I_l , {I_m}^{\dag}] = - T_{l+m},  
\eeq
whence,
\beq
[I_l , H - \mu \hat{N} ] = -2 {\mu} I_l + 2 I_{l+1} -  U_s T_l I_0. 
\eeq
Now, we invoke the  law of large numbers  and
argue that in the present problem the operator product $T_l I_0 $ can
be replaced by $ \mcl \mu_l I_0$ where $ \mcl \; \mu_l~=~< ~T_l~> $ ,
the thermodynamic average of $T_l$, which is clearly of order $\mcl$
\cite {er_fac}. Then
the equations of motion for the usual time ordered Green functions
$<<I_l;{I_m}^{\dag}>> $ reduce to the closed set
\beq
( i\Omega_n + 2 \mu) <<I_l;{I_m}^{\dag}>> \; = \; -\mcl \mu_{l+m} + 
2 <<I_{l+1};{I_m}^{\dag}>> - \mcl U_s {\mu}_l <<I_0;{I_m}^{\dag}>>
\label{factored}
\eeq
 
which can be solved exactly. The solution is given by 
 
\beq
<<I_l;{I_m}^{\dag}>> \; = \; - \mcl \Pi_{l+m} \; + 
 \; \mcl {\Pi_l} \;\frac {U_s} {1 + U_s \mcl {\Pi_0} } \; \mcl {\Pi_m }
\label{eqmsol}
\eeq
where 
$$\Pi_l (i\Omega_n) 
\equiv \sum_{l'=0}^{\infty} \frac { 2^{l'} \mu_ {l+l'} }{{( i\Omega_n + 2
\mu)}^{l'+1} }  =  (1/\mcl) \sum_k \frac {(\epsilon_k)^l (2 <n_k>
-1)}{(i\Omega_n - 2 \xi_k)}. $$ 
This may be verified using the easily derived
recursion relation $ \Pi_{l+1} =[(i\Omega_n + 2 \mu) \Pi_l - \mu_l]/2 $.
Furthermore, one of the key features of the models being discussed is
that {\em the one particle propagators are unrenormalised } \cite{g_ur}, 
so that $<n_k>=f(k)$ whence, $ \Pi_l (i\Omega_n) = (1/ \mcl)\sum_{k}
{(\epsilon_k)}^l {\pi_{k,n}} $. Then it is easily seen that the results in 
Eq. (\ref{eqmsol}) are basically the same as in  Eq. (\ref
{main}). Starting from the latter, multiplying by  $({\epsilon_k})^l
({\epsilon_{k'}})^m $ and summing over $k$ and $k'$, we get the former.

An analysis of the detailed properties of the function $V_{eff}$ is of
crucial importance for the rest of our discussion. For a simple model
bandstructure, with a  constant density of states $g(\epsilon)=1/2$ for
$-1 < \epsilon < 1$, we can compute it exactly as
\beq
V_{eff} (i\Omega) \rightarrow V_{eff}(\omega + i \eta)
=    \frac{V}{1 + V \Pi_0(\omega)} ,
\eeq
with
$$
\Pi_0(\omega) = \frac{1}{4}
\log( \frac{ 4 |(\omega/2 + \mu)^2 -1 |} { \omega^2} ) + i \frac{\pi}{4}
\theta(2 - |2 \mu + \omega |) sgn (\omega)
$$
The band extends from $-2 + 2 \delta$ to $ 2 + 2
\delta$, where the hole filling $\delta = 1 - \rho = -\mu$. 
The schematic behaviour of  $V_{eff}$ is 
as follows. At very high  frequencies, $V_{eff} \approx V $. It has two
poles at certain  large frequencies
that essentially dominate its physics . In between these
poles lies the intermediate frequency range where $V_{eff}$ is of
order unity, which contains the branch cut 
corresponding to the one electron band of states. The  location of the 
poles can be found  by using a large frequency expansion for the function
 $\Pi_0 (i\Omega_n)$, which for $|\Omega_n| >> W $ ( $W$  is the band
width) behaves as 

\beq
 \Pi_0 (i\Omega_n)= -\frac{\delta}{(i\Omega_n)} -  \frac{ 2 (
|\mu_1| - \mu \delta)}{(i \Omega_n)^2} + O(\frac{1}{|\Omega_n|^3}). 
\label{pilarge} 
\eeq
The higher order terms can be verified to be negligible. The poles of the
$V_{eff}$ can then be found by solving for the zeroes of the denominator,
a quadratic in $\zeta \equiv 1/ i \Omega_n$ , given as
$0= V^{-1} - \delta \zeta - 2 ( |\mu_1| - \mu \delta) \zeta^2 $. The roots
are always real corresponding to real frequency poles of $V_{eff}$. 
Hence,(a) at half filling, $\delta=0$, the poles are at 
$\pm \omega_0 = \pm \sqrt{2 V |\mu_1|}$, and 
near the poles $V_{eff} \approx \pm \frac{\omega_0 V}{2( \omega \mp
\omega_0)}$. The pole frequencies are
infinite in  the thermodynamic limit, but leave behind consequences in the
groundstate as we see later. (b) away from but close to half filling ,
$\delta << 1$, one pole  is at  $\omega_2
\equiv V \delta$ with residue $V^2 \delta$, and the other at a large
negative frequency $- \omega_1 = -(2 |\mu_1| / \delta - 2\mu) $ with
residue $ -\omega_1^2/\delta$. As $\delta \rightarrow 0 $, the
latter poles smoothly go over into those of the first case. The
contribution from the branch cut does not have any particularly simple
form, but is not  important  in the most interesting region of the
problem, namely $\delta \sim 0$. 

Using the above properties of $V_{eff}$ we can compute exactly the
instantaneous (expectation) values of the the extended s-wave
and the s-wave correlation functions :\\
$\alpha \equiv \frac{1}{4 \mcl} 
\; < A^\dagger A>$  with $A = [T,B] = -2 \sum_k \epsilon_k b_k$, and 
${\cbeta}   \equiv \frac{1}{ \mcl} \; < B^\dagger B>$. \\
Using the main result Eq(\ref{main}), we find

\begin{eqnarray}
\alpha 
&=& k_B T \sum_n  [ \Pi_2 (i\Omega_n) - V_{eff} (i\Omega_n)\Pi_1^2
(i\Omega_n) ] e^ {(i\Omega_n 0^+)} \nonumber \\  
{ \cbeta} 
 & = & \frac{ k_B T}{V^2}
\sum_n    [ V - V_{eff} (i\Omega_n)] e^ {(i \Omega_n 0^+)}.  \label{beta}
\end{eqnarray}

The main  contributions to the resulting frequency sums are quite easily
seen to be tied to the  pole contributions of $V_{eff}$, the branch cut
part giving an uninteresting subleading contribution. 
We find at half filling  
\beq
\alpha_{pole} =  \frac{ \sqrt{ U_s \mcl | \mu_1|^3 }} { 2\sqrt{2} } \; ; \;
\cbeta_{pole} =  \frac{\sqrt{|\mu_1|}}{\sqrt{2 U_s \mcl}}.
\label{betares1} 
\eeq
Near half filling we find
\beq
\alpha_{pole} = \frac{\mu_1^2}{\delta} \; ; \;
{\cbeta}_{pole} = \frac{4 \mu_1^2}{\delta^3 U_s^2 \mcl^2}.  
\label{betares2}
\eeq
At half filling Eq.(\ref{betares1}) gives us the quasi long
ranged order as well as the correction to ground state energy along with
their appropriate coeffecients, the latter by an integration over the
coupling constant $U_s$. These answers are in good numerical agreement
with exact numerics on a finite sized system \cite{bss3}. Away from half
filling we find that the uncertainty principle lower bounds are off
by a factor of 2, i.e., from Eq.(\ref{betares2})  
$\alpha_{pole} = 2 \alpha_{Lower Bound}$.

The results given above enable us to compute several
other response functions exactly.  For example we find that the charge
stiffness or Meissner fraction is non- zero, and in fact unchanged from
the non interacting value at half filling. We also find at half filling
that  the appropriate order parameter density  $\Delta(r)$ has
correlations that are  novel  : $< \Delta^\dagger (r) \Delta(0)> \sim
\frac{c}{\sqrt{{\mcl}}}$. Thus although we do not have LRO of the
usual sort, one has a divergent ``structure function'' $\int \; d\vec{
r}\,< \Delta^\dagger(r) \Delta(0)> $.  

We next discuss the important
case of  $\zeta(k) = \cos(k_x) - \cos(k_y)$ in two dimensions.
 This corresponds to suppressing d-wave order at the length scale 
of nearest neighbours.From the uncertainty principle argument of  Refs
\cite{bss1,bss2}, it follows that this would lead to enhancement again in
the d-wave channel, but at a  longer length scale, i.e. the resulting
$\hat{A} = - 2 \sum \epsilon_k \zeta(k) b_k$ should have enhanced
correlations. The functional integral solution sketched here bears this
out exactly. We recover the results in Eq.s(\ref{betares1},\ref{betares2})
with the replacements : 
$\delta \rightarrow \hat{\delta} =   {\mcl}^{-1} \sum_k \zeta^2(k) ( 1 -
2 f(k))$  and  $\mu \rightarrow \hat{\mu}_1= {\mcl}^{-1} \sum_k  
\epsilon_k \zeta^2(k)( 1 - 2 f(k))  $. The enhanced
correlation  function of $\hat{A}$ then diverges at the point where
$\hat{\delta}$ vanishes.  So long as the one electron dispersion has the
bipartite symmetry, one can see that $\hat{\delta}$ vanishes exactly at
half filling. However, if the dispersion does not have this symmetry, e.g.
by having a second neighbouring hopping $t'$, then $\hat{\delta}$ vanishes
at some other density determined by $t'$, as illustrated in Fig.1. The
case of $t' \sim -.4$ is popular in literature, since it leads to a Fermi
surface that is consistent with that seen in the photo-emission
experiments in High $T_c$ systems \cite{pes}, and it is an amusing
coincidence that the filling $\delta \sim .18 $ is close to the optimum
doping . 

Finally, we can show  that many of the above features of the
infinite range model retain their relevance for more realistic
models with finite range repulsion. To see this, consider (in the s wave
case) a more general  model of the form $H = T + \frac{1}{\mcl} \sum U(p)
\; B^\dagger(p) B(p)$,  where $B(p)= \sum_k c_{-k, \downarrow} c_{k+p,
\uparrow}$ is a Cooper pair operator  with total momentum $p$ , with 
$U(p) = U_s (l_c)^{d} $ for a small set of  $\mcl / (l_c)^{d} $ points
surrounding the centre of the Brillouin zone. This would correspond to  
repulsive interactions with a long but finite range  $l_c$ \cite{fnoteham}.
Then it is not hard to see that   the results we have discussed above would
retain much of their validity (but for some differences in details) to
leading order in $(1/l_c)$, with the replacement $ U_s \mcl  \rightarrow
U_s (l_c)^{d} $, leading to enhanced pairing correlations over a finite
range in momentum space. The quasi Long Ranged Order would be replaced by
an enhancement of the $O(1)$, like that in the single mode model away from
half filling. In this case, we see that all Cooper pairs with a
finite ( but small) center of mass momentum are also influenced by the
interaction, and thus the model is more realistic by way of helping
current carrying states.

In summary, we have found an exact solution for an interesting model of
Fermions with purely repulsive interactions with infinite range, which 
may be regarded as a  {\em meanfield repulsive
model}. The resulting solution has quasi long ranged order at half  
filling,  as well as  large unbounded enhancements as one approaches half
filling, in the equal time pairing correlations. We have also argued
that the above methods and results retain their relevance even
for generalised models where the repulsive interactions have a large but
finite range, and are therefore more realistic.  Finally, it
is remarkable  that the enhanced pairing correlations in these models arise
from very high energy poles in the scattering amplitude, not unlike the
physics of Mott Hubbard systems, where the upper Hubbard band influences
the properties of carriers in the lower band.

One of us (BSS) is grateful to the Aspen Center of Physics for  support,
where a  part of this paper was written. He thanks P W Anderson and B
Shraiman for  stimulating discussions and   constructive suggestions.

\begin{figure}
  \begin{center}
  \leavevmode
  \epsfxsize=9.5cm
  \epsfbox{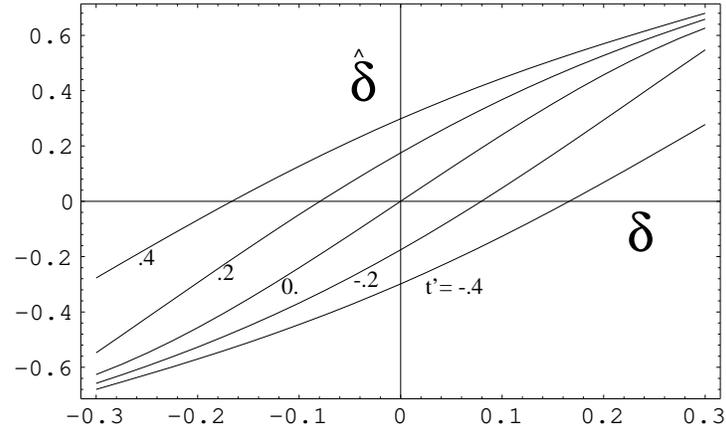}
  \end{center}
  \caption{ The effective filling  {$ \hat{\delta}$} for d-wave 
superconductivity against the true filling { $\delta $ } for different
values of the second neighbour hopping $t'$.      
\label{fig3}}
\end{figure}
*{\it Also at the JNCASR, Jakkur, Bangalore }


\begin{thebibliography}{99}
\bibitem{posit} This claim  rests upon the positivity  of the real part
of the coefficient of the quadratic term, for all frequencies, which is
easily verified by the explicit  form given above.
\bibitem{fnotecorr} If one retained corrections to the next order in
$(1/ \mcl) $, the denominator in Eqs(\ref{basic},\ref{veffective}) would
read $1 + U_s \mcl \Pi_0(i \Omega_n) + \gamma( i \Omega_n)$, where
$\gamma= U_s^2 \mcl \sum_m \Gamma_{(m,n,m,n)} < \phi^*(m) \phi(m) >$ .
We can show from the explicit form of $\Gamma$  that such corrections 
are completely negligible as  regards the high frequency spectral 
features of $V_{eff}$ relevant for the rest of this paper.
\bibitem{er_fac}  The error involved in the factorization is clearly
subleading in $\mcl $. From the
solution Eq. (\ref{eqmsol}) which implies that $<<I_0;I^\dagger_m >> = -
\mcl \Pi_m /( 1 + \mcl U_s \Pi_0)$, and hence is of $O(1)$, we see that .
the omitted term in Eq(\ref{factored}) is actually non extensive. 
For related comments, see \cite{fnotecorr} above.
\bibitem{g_ur}  This can be shown by observing that  our  $V_{eff}
(i\Omega_n)$ in Eq(\ref{veffective}) is the exact one   particle
irreducible (1PI) vertex function for the model given by  Eq.
(\ref{hamiltonian}), (with  $i\Omega_n $, the frequency in the pairing
channel, being appropriately related to the frequencies of the external
legs, and the momentum in the pairing cannel being constrained to vanish) 
and using this 1PI vertex function to calculate the self energy exactly.
\end{thebibliography}
\end{document}